\documentclass[nofootinbib,onecolumn,aps,preprintnumbers]{revtex4-2}
\usepackage{epsfig,graphics}
\usepackage{amsfonts,amssymb,amsmath}            
\usepackage{ifthen}
\usepackage{amsthm}
\usepackage{hyperref}
\usepackage{url}
\usepackage{todonotes}
\usepackage{color, colortbl,hhline}
\usepackage{tikz}
\usetikzlibrary{matrix}
\usepackage{diagbox}
\usepackage{algorithm}
\usepackage{algpseudocode}

\newcommand{\comment}[1]{}
\newcommand{\ket}[1]{| #1 \rangle}
\newcommand{\bra}[1]{\langle #1 |}

\newcommand{\tr}{{\rm tr}}

\newcommand{\bs}[1]{\boldsymbol{#1}}

\theoremstyle{plain}
\newtheorem{theorem}{Theorem}
\newtheorem{proposition}[theorem]{Proposition}
\newtheorem{lemma}[theorem]{Lemma}

\theoremstyle{definition}

\definecolor{amber}{rgb}{1.0, 0.75, 0.0}
\definecolor{aureolin}{rgb}{0.99, 0.93, 0.0}
\definecolor{steel}{HTML}{91CDE2}
\definecolor{bblue}{HTML}{508AA8}
\definecolor{beyes}{HTML}{EDF6FC}

\newcommand{\coker}{\mathrm{coker}}
\newcommand{\Tr}{\mathrm{Tr}}

\begin{document}
	
	\title{Extremal Steering Assemblages}
	\date{\today}
	
	\author{Thomas \surname{Cope}}
	\email{thomas.cope@itp.uni-hannover.de}
	\address{Institut f{\"u}r Theoretische Physik, Leibniz Universit{\"a}t Hannover, Appelstr. 2, 30167 Hannover, Germany}
	\author{Tobias J. \surname{Osborne}}
	\email{tobias.osborne@itp.uni-hannover.de}
	\address{Institut f{\"u}r Theoretische Physik, Leibniz Universit{\"a}t Hannover, Appelstr. 2, 30167 Hannover, Germany}
	
	\begin{abstract}
Non-local correlations between a fully characterised quantum system and an untrusted black box device are described by an assemblage of conditional quantum states. These assemblages form a convex set, whose extremal points are relevant in many operational contexts. We give necessary and sufficient conditions for an assemblage to be extremal using linear independence conditions, and an algorithm to decompose a generic assemblage into extremal points. A Matlab implementation of this algorithm is provided in the supplementary material. 
	\end{abstract}
	\maketitle
	
\section{Introduction}
The concept of steering has been part of quantum mechanics since its inception. It played a key role in the argument put forward by Einstein, Podolsky and Rosen, whose 1935 paper \cite{EPR1935} exposed the ability of quantum systems to remotely influence one another, even when distantly separated. The phenomenon of steering was explored further by Schr{\"o}dinger in two papers \cite{S1935,S1936}, who characterised exactly how this influence occurs.

Steering is currently a topic of considerable interest, as it is the resource used in \emph{semi} Device Independent Quantum Key Distribution (DIQKD) \cite{BCWSW2012,WBLZL2013,GA2015,KWW2020}. These protocols provide a trade-off between the security of DIQKD \cite{MY1998,BHK2005,AGM2006,ADFRV2018} and the high key rates of Quantum Key Distribution (QKD) \cite{BB1984,E1991,R1999,PMLB2008,Pirandolaetal2020}. 

In applications, steering is not usually described as a property of quantum states, but of \emph{assemblages} \cite{WJD2007}: the collection of possible states a subsystem may be steered into. In this paper, we characterise exactly when an assemblage is extremal, within the convex set of all assemblages. Extremality is an important property of assemblages since several steering measures are convex, such as the steering weight \cite{SNC2014}, steering robustness \cite{PW2015}, and relative entropy of steering \cite{GA2015,KW2017}. Furthermore, extremality allows for an easy calculation of an assemblage's entanglement of formation \cite{C2021}.

\section{Preliminaries}
We begin with the definition of an assemblage. An assemblage is a collection of sub-normalised density operators $\{\sigma_{n|r}\}$, which satisfy the conditions 
\begin{align*}
	\sigma_{n|r}&\geq 0\;\;\forall n,r, &&\mathrm{and}& \sum_{n} \sigma_{n|r}&= \sum_{n} \bs{\sigma}_{n|r'} \;\forall r,r',
\end{align*}
known as the positivity and no-signalling conditions respectively. The state $\rho_{B}:=\sum_{n} \sigma_{n|r}$ is often referred to as the \emph{marginal} state. We consider the set of all such possible collections as our general set of steering assemblages $\mathcal{G}$. This definition arises from the result of two spatially separated parties sharing a quantum state, $\rho_{AB}$, with each substate representing the probabilistic state obtained from a choice of measurement (labelled $r$) on one subsystem, given by $\sigma_{n|r}=\mathrm{Tr}_{A}\left[\left(M_{n|r}\otimes \mathbb{I}_{B}\right)\rho_{AB}\right]$. However, we do not assume knowledge of the original state or measurements, studying only the resulting set of substates.\\

The set of assemblages is convex and as such, any point can be decomposed into a convex combination of extremal points. A point $\mathbf{x}$ in a convex set is called extremal if any convex decomposition of the form $\mathbf{x}=p\mathbf{x}_1+(1-p)\mathbf{x}_2$, with $p\in(0,1)$, implies $\mathbf{x}_1=\mathbf{x}_2=\mathbf{x}$ (i.e. there are no non-trivial convex decompositions of $\mathbf{x}$).

When discussing assemblages it is often useful to consider the case where we restrict the sets $n\in\{1\ldots N\}$, $r\in\{1\ldots R\}$ and the dimension of our Hilbert space is fixed. We write this set as $\mathcal{G}_{N|R}^{d}$, and express assemblages within it as an ordered tuple $\bs{\sigma}=\left(\sigma_{1|1}\ldots \sigma_{N|1}\ldots \sigma_{1|R}\ldots \sigma_{N|R}\right)$. In this paper, we show when an assemblage is extremal in the restricted set $G_{N|R}^{d}$. However, this extremality holds for the general set $\mathcal{G}$, since they may be embedded into the general set by adding either trivial outcomes ( $\sigma_{n|r}=0,\; \forall n>N$ ), trivial dimensions (  $\mathrm{dim}( \cup \mathrm{coker}(\sigma_{n|r}))\leq d$ ), or by repeating inputs ( $\sigma_{n|r}=\sigma_{n|r'},\,\forall r>R$ ). All of these embeddings preserve extremality.

Any bipartite assemblage with marginal $\rho_{B}=\sum_{i}\lambda_{i}\ket{i}\bra{i}$ may be created via a specific choice of quantum state and measurements: by taking the pure state defined by the purification $\ket{\Phi}=\sum_{i}\sqrt{\lambda}_{i}\ket{i}_{A}\otimes\ket{i}_{B}$, and measurement operators $M_{n|r}=\rho_{B}^{-1/2}\sigma_{n|r}^{T}\rho_{B}^{-1/2}$. The transposition is with respect to the basis $\{\ket{i}\}$. However, this is generally not the only state and measurements that reproduce a given assemblage.  \\

The above construction connects many properties of assemblages to those of sets of Positive Operator Valued Measures (POVMs). Our work in this paper is motivated by \cite{APP2005}, who studied the extremality of POVMs. However, we stress that extremality of an assemblage is not trivially derived by its corresponding POVMs. A simple example of this can be seen by the assemblage $\sigma_{0|0}=\ket{0}\bra{0}/2, \sigma_{1|0}=\ket{1}\bra{1}/2$. This assemblage can be decomposed with equal weight into two assemblages $\sigma^0_{i|0}=\delta_{i0}\ket{0}\bra{0}$ and $\sigma^1_{i|0}=\delta_{i1}\ket{1}\bra{1}$. However, the corresponding POVM, $\{\ket{0}\bra{0},\ket{1}\bra{1}\}$ cannot be decomposed.\\

\section{Extremal Assemblages}

Let us now begin with the characterisation of extremal points of the set $\mathcal{G}_{N|R}^{d}$. Let us consider a convex decomposition of an assemblage, of the form $\bs{\sigma}=p\bs{\sigma}^1+(1-p)\bs{\sigma}^2$. The two decomposing points must themselves be assemblages and satisfy the no-signalling condition.
In this paper we follow the method of \cite{APP2005}, who explore extremality using the idea of perturbations.  Let us suppose we have two valid assemblages of the form $\bs{\sigma}^{\pm}=\bs{\sigma}\pm\bs{D}$, where $\bs{D}$ is some perturbation vector. The non-existence of a $\bs{D}\neq \bf{0}$, such that the defined $\bs{\sigma}^{\pm}$ are valid assemblages, is a necessary and sufficient condition for $\bs{\sigma}$ to be extremal. Neccessity comes from the fact that $\bs{\sigma}=1/2(\bs{\sigma}^{+}+\bs{\sigma}^{-})$. To show sufficiency, suppose we have a valid non-trivial decomposition $\bs{\sigma}=p\bs{\sigma}^1 +(1-p)\bs{\sigma}^2$, and assume w.l.o.g. that $p\geq 1/2$. Then by defining $\boldsymbol{D}=\bs{\sigma}^1-\bs{\sigma}$, we can rewrite $\bs{\sigma}=1/2(\bs{\sigma}+\boldsymbol{D})+1/2(\bs{\sigma}-\boldsymbol{D})$.
As $\bs{\sigma}+\boldsymbol{D}=\bs{\sigma}^1$, we need only check that $\bs{\sigma}-\boldsymbol{D}$ is a valid assemblage. However, we can do this by substituting in the convex decomposition of $\bs{\sigma}$:
\begin{equation}
\bs{\sigma}-\boldsymbol{D} = p\bs{\sigma}^1+(1-p)\bs{\sigma}^2 - \bs{\sigma}^1 +  p\bs{\sigma}^1+(1-p)\bs{\sigma}^2 = (2p-1)\bs{\sigma}^1 +(2-2p)\bs{\sigma}^2
\end{equation}
Since we assumed $1/2\leq p \leq 1$, both $(2p-1),(2-2p)\geq 0$ and add up to 1. This means $\bs{\sigma}-\boldsymbol{D}$ is a valid convex combination of two assemblages, and thus itself a valid assemblage. \\

Let us consider what properties $\boldsymbol{D}$ must therefore have. First we require every element of $\bs{\sigma}-\bs{D}$ to be positive, implying that $\mathrm{ker}(\sigma_{n|r})\subseteq \mathrm{ker}(D_{n|r})$, or equivalently $\coker(D_{n|r})\subseteq \coker(\sigma_{n|r})$. Secondly, we must have $\Tr(\sum_{n}\sigma_{n|r} \pm D_{n|r})=\Tr(\sum_{n}\sigma_{n|r}) =1$, and thus by linearity $\Tr(\sum_{n}D_{n|r})=0$. Thirdly, as as both $\bs{\sigma}^1, \bs{\sigma}^2$ satisfy the no-signalling condition, so too must the difference between them, $\bs{D}$. Finally, every $D_{n|r}$ is the difference between two Hermitian operators, so they too must be Hermitian. These are in fact the only constraints that must hold for $\bs{D}$, and so we arrive at the first statement of extremality.\\

\begin{theorem}
An assemblage $\bs{\sigma}$ is extremal iff there exist no non-trivial solutions to the equations:
\begin{align}
\sum_{n} D_{n|r} &= \sum_{n} D_{n|r'},\; \forall r,r', & \Tr\left(\sum_n D_{n|r}\right) &= 0, & \coker(D_{n|r})&\subseteq \coker(\sigma_{n|r})\; \forall n,r
\end{align}
\end{theorem}

where the $D_{n|r}$ are Hermitian. It is useful to rewrite the above result using an explicit basis for each $D_{n|r}$. To do this, for each $\sigma_{n|r}$ we define $\{\ket{v_{n|r}^a}\}_{a}$ as the set of eigenvectors of $\sigma_{n|r}$ which correspond to non-zero eigenvalues of $\sigma_{n|r}$. We then represent $D_{n|r}$ in this basis. Our above statement then becomes: $\bs{\sigma}$ is extremal iff
\begin{align}
	&\sum_{n}\sum^{\mathrm{rank}(\sigma_{n|r})}_{a,b=1} D^{a,b}_{n|r}\ket{v_{n|r}^a}\bra{v_{n|r}^b} = \sum_{n}\sum^{\mathrm{rank}(\sigma_{n|r'})}_{a',b'=1}D^{a',b'}_{n|r'}\ket{v_{n|r'}^{a'}}\bra{v_{n|r'}^{b'}} ,\; \forall r,r', \text{   and  }
	\sum_{n}\sum_{a} D^{a,a}_{n|r}=0,\\
\Rightarrow &\;\;D^{a,b}_{n|r}=0\;\; \forall a,b,n,r
\end{align}

with the underlying requirement that $D^{a,b}_{n|r}=\bar{D}^{b,a}_{n|r}$. We can recognise this as a set of \emph{linear independence} conditions, for the operators $\{\ket{v_{n|r}^a}\bra{v_{n|r}^b}\}$. 

Before we look at these conditions in further detail, it is worth noting what happens in the special case where $R=1$. Here we have no no-signalling constraints, reducing our condition to:
\begin{equation}
\sum_{n}\sum_{a} D^{a,a}_{n|r}=0, 
\Rightarrow D^{a,b}_{n|r}=0 \;\;\forall a,b,n,r.
\end{equation}
Such an implication can only be true in one specific instance: when $n,a$ can only take one possible value. 
This means we have the following result: 
\begin{lemma}\label{lemmaP}
All extremal assemblages with $R=1$ are of the form
$\sigma_{n|1}=\delta_{nn'}\ket{\phi}\bra{\phi}$.
\end{lemma}

In the case of $R>1$, things are more complicated. Note that the no-signalling conditions are operator constraints, whilst the trace condition and Hermicity requirements are coefficient constraints. We can therefore incorporate these constraints into our operator constraints, by modifying the operators involved. This leads us to another statement of extremality: 
\begin{proposition}\label{FullTheorem}
An assemblage $\sigma_{n|r}$ is extremal iff $\forall$ pairs  $r,r'$, the following operators are linearly independent:

\begin{align}
		&\frac{1}{\sqrt{2}}\left(\ket{v_{n|\tilde{r}}^{a}}\bra{v_{n|\tilde{r}}^{a}}-\ket{v_{1|\tilde{r}}^{1}}\bra{v_{1|\tilde{r}}^{1}}\right)\label{fullcheck},&
		&\frac{1}{\sqrt{2}}\left(\ket{v_{n|\tilde{r}}^{a}}\bra{v_{n|\tilde{r}}^{b}}+\ket{v_{n|\tilde{r}}^{b}}\bra{v_{n|\tilde{r}}^{a}}\right),& &\frac{1}{\sqrt{2}i}\left(\ket{v_{n|\tilde{r}}^{a}}\bra{v_{n|\tilde{r}}^{b}}-\ket{v_{n|\tilde{r}}^{b}}\bra{v_{n|\tilde{r}}^{a}}\right),
\end{align}
where $n$ takes any value, $a,b$ are limited by the rank of $\sigma_{n|r}$ and $\tilde{r}$ ranges through both $\{r,r'\}$. The trivial operator obtained when $n=1,a=1$ is excluded.
\end{proposition}
The linear dependence here refers to \emph{real coefficients}, since finite Hermitian operators form a real vector space. The above statement is complete, in that it is a necessary and sufficient condition for the extremality of a given assemblage. To do so 
 requires one to check $r(r-1)/2$ linear independence conditions; requiring a quadratic (in inputs) number  of checks to prove extremality. We can often simplify this number by splitting linear dependence into two types: zero-marginal and non-zero-marginal. A zero-marginal linear dependence is of the form:
\begin{equation}
	\sum_{n}\sum^{\mathrm{rank}(\sigma_{n|r})}_{a,b=1} D^{a,b}_{n|r}\ket{v_{n|r}^a}\bra{v_{n|r}^b} =0,\;\; \exists\, D^{a,b}_{n|r}\neq 0.\label{zeromargin}
\end{equation}
Note that this, along with setting $D^{a,b}_{n|r'}=0,\;\forall r'\neq r$, gives us a valid non-trivial perturbation, proving $\bs{\sigma}$ is non-extremal. As the name implies, the marginal of $\mathbf{D}$ is 0.\\
By contrast a non-zero-marginal perturbation is of the form:
\begin{equation}
	\sum_{n}\sum^{\mathrm{rank}(\sigma_{n|r})}_{a,b=1} D^{a,b}_{n|r}\ket{v_{n|r}^a}\bra{v_{n|r}^b} = X_{D}\neq 0,\;\forall r,\; \tr(X_{D})=0.
\end{equation}
The advantage of this distinction is that Eq. (\ref{zeromargin}) can be checked for each $r$ individually, with every check involving a smaller number of operators than the conditions in Eq. (\ref{fullcheck}). If all of these checks are passed, then we turn to the possibility of non-zero-marginal perturbations; and we must check the linear independence conditions of Eq. (\ref{fullcheck}) for one pair of inputs $r,r'$. If this test is passed then we can conclude the assemblage is extremal! This is because the linear independence of these operators forces any marginal $X_{D}$ to be equal to the zero operator; however, we have shown with our previous checks that such a perturbation is impossible. This gives another equivalent condition of extremality.
\begin{proposition}
$\mathbf{\sigma}$ is extremal iff the operators:
\begin{align}
&\ket{v_{n|r}^{a}}\bra{v_{n|r}^{a}},&
&\frac{1}{\sqrt{2}}\left(\ket{v_{n|r}^{a}}\bra{v_{n|r}^{b}}+\ket{v_{n|r}^{b}}\bra{v_{n|r}^{a}}\right),& &\frac{1}{\sqrt{2}i}\left(\ket{v_{n|r}^{a}}\bra{v_{n|r}^{b}}-\ket{v_{n|r}^{b}}\bra{v_{n|r}^{a}}\right),& \;\;&\forall (n,a,b) \; a,b\leq \mathrm{rank}(\sigma_{n|r})
\end{align}
are linearly independent for all $r$ individually,
and the operators defined in proposition \ref{FullTheorem} are linearly independent for any pair $r,r'$.\\
\end{proposition}
Note that if a pair $r,r'$ exhibit linear dependence, we cannot conclude that the assemblage is \emph{not} extremal; we have to keep checking until a $r,r'$ pair that passes the check is found. This means that in the worst case (the only linearly independent pair is found last) just checking all the pairs may be faster. 

\section{An algorithm to decompose into extremal assemblages}
Together with this paper we provide an algorithm to decompose an arbitrary assemblage into extremal ones. The pseudo-code can be found in the supplementary material.\\

Our implementation uses proposition 4 as the criterion of extremality, and is written to be recursive, since after every perturbation has been applied, there are two more assemblages whose extremality needs to be checked. This means the number of calls grows exponentially with the number of possible perturbations. Despite this it remains viable for decomposing low-dimensional assemblages, with its performance greatly improved when used only to test extremality, rather than to find the full decomposition. A Matlab implementation is also provided.\\

We now provide a few examples of decomposed assemblages to illustrate the working of the algorithm. All examples can also be found in the supplementary material.\\

\textbf{Example 1}: We take our first example from \cite{SGBD2013}. They consider a POVM whose effects are given by $\Pi_{a} = 2/5\ket{\phi_{a}}\bra{\phi_{a}}$, where $\ket{\phi_{a}}$ is the qubit state with Bloch vector $(\cos(2\pi a/5),\sin(2\pi a/5),0)$. This measurement can be decomposed into extremal ``trines"; POVMs with three non-trivial effects.\\
We can adapt this example into an assemblage by defining $\sigma_{a}=\Pi_{a}/2$; so that the marginal now gives $\mathbb{I}_2/2$, a valid density operator. Decomposing this assemblage into extremals, we find that $\bs{\sigma}=\sum_{i}1/5\bs{\sigma}^{i}$, where 
$\sigma_{a}^{i}=\delta_{ia}\ket{\phi_{a}}\bra{\phi_{a}}$. This fits with what we expected from lemma \ref{lemmaP}, as  $\bs{\sigma}$ is an assemblage with one input choice. This example highlights how POVM and assemblages decompositions are not analogous, since one is unable to form a qubit POVM from a single $\bs{\sigma}^{i}$ (as it spans a 1-dimensional subspace).\\

\textbf{Example 2}: In our second example we take the maximally entangled qubit $\ket{\Phi^+}=(\ket{00}+\ket{11})/\sqrt{2}$, and create as assemblage using two measurements: a projective measurement in the Pauli-X basis; and a POVM whose effects are $\Pi_{a}=1/2\ket{\phi_{a}}\bra{\phi_{a}}$, where the set of Bloch vectors for $\ket{\phi_{a}}$ are $\{(0,0,1),(\sqrt{8/9},0,-1/3),(-\sqrt{2/9},\sqrt{2/3},-1/3),$\\$(-\sqrt{2/9},-\sqrt{2/3},-1/3)\}$, forming a regular tetrahedron. Since this assemblage is formed by measuring an extremal quantum state with two extremal POVMs, one might expect the assemblage itself to also be extremal. However, the assemblage is decomposable $\bs{\sigma}=2/3\bs{\sigma}^{1}+1/3\bs{\sigma}^{2}$. The two assemblages can be characterised by their marginals, whose Bloch vectors are the points $\mathbf{v}_1=(0,0,-1/2\sqrt{2})$; $\mathbf{v}_2=(0,0,1/\sqrt{2})$. These are exactly the points where the tetrahedron intersects with the $x$-axis. \\

\textbf{Example 3}:
In this example we take the assemblage obtained from measuring a qutrit maximally entangled state with two Mutually Unbiased Measurements (MUBs). Such assemblages appear in the study of dimension verification \cite{DSUVMMB2021}. We can then mix this assemblage with white noise; an assemblage whose substates are identically proportional to the identity. By definition this is not an extremal assemblage. With decomposition via the algorithm, we may decompose $4/5\bs{\sigma}^{\mathrm{MUB}}+1/5\bs{\sigma}^{WN}$ into 3727 inequivalent assemblages, none of which are $\bs{\sigma}^{\mathrm{MUB}}$ (which can be checked to be extremal). This large number should serve to remind us that the perturbation method does not help to find a minimal decomposition. The full decomposition is provided in the supplementary material.

\section{Discussion and Conclusions}
In this paper, we have given several equivalent definitions which are necessary and sufficient for an assemblage to be extremal. These can be directly tested by calculating linear dependencies of vectors created from the eigenvectors of the assemblage's substates. We also provide an algorithm which can decompose an assemblage into extremal assemblages, and provide a Matlab implementation. This algorithm scales poorly with the dimension and number of inputs of the assemblage; this is because every perturbation creates two new assemblages which must then themselves be decomposed, resulting in exponential behaviour. In \cite{SGBD2013} a much more efficient method for decomposition of POVMs was presented, exploiting linear programming. However, we were unable to transfer the concept to steering assemblages.

We also comment that our conditions provide a sufficient condition for extremality of a multipartite steering assemblage. This is because a multipartite assemblage (with two spatially separated measuring parties) may be thought of as a restricted case of the bipartite scenario, where Alice's outcomes satisfy an additional no-signalling constraint. One may construct a counter-example to necessity by considering a non-signalling probability distribution, which may be decomposed into signalling components (since in the bipartite case, no such restrictions on Alice's side apply) e.g. the PR-Box example discussed in \cite{SBCSV2015}.\\

Many questions about steering assemblages can be phrased instead as questions regarding Positive Operator Valued Measurements (POVMs) via the mapping $\sigma_{n|r}\rightarrow \rho_{B}^{-1/2}\sigma_{n|r}\rho_{B}^{-1/2}$. Note that extremality of the resultant POVMs does not imply extremality of the original assemblage: the requirement of POVM elements to sum to the identity necessarily preserves the marginal.\\

Given that assemblages may be created from a given state $\rho$ and set of measurements $\{M_{n|r}\}$ via the formula $\sigma_{n|r}=\mathrm{Tr}\left[\left(M_{n|r}\otimes \mathbb{I}_{B}\right)\rho_{AB}\right]$, it is natural to wonder how convex decompositions of the state or measurements translate into decompositions of the assemblage. From this formula, one notes that a decomposition of $\{M_{n|r}\}$ for a given $r$ results in a possible marginal-invariant perturbation, while in general decompositions of the state lead to perturbations which change the marginal. However, one should not draw the tempting conclusion that marginal-preserving perturbations \emph{must} arise from measurement decompositions: by taking $\rho=p\ket{\Phi^+}\bra{\Phi^+} +(1-p)\left(\ket{00}\bra{00}/2+\ket{11}\bra{11}/2\right)$, $M_{1|1}=\ket{0}\bra{0},M_{2|1}=\ket{1}\bra{1},M_{1|2}=\ket{+}\bra{+},M_{2|2}=\ket{-}\bra{-}$, one creates an assemblage with a non-trivial perturbation leaving both $\sigma_{i|1}$ invariant, therefore preserving the marginal.\\

One potential application of this work may be in the field of self-testing: in which one can verify the underlying states and measurements via the observed assemblage. For quantum correlation self-testing of both states and measurements, it has been shown the required point must be extremal and finite-dimensional realisable \cite{Gohetal2018}. It would be interesting to try and connect this with steering based self-testing \cite{SH2016,GBDSJM2018}. Another application may be in the semi-device independent certification of entanglement, which has been investigated in \cite{CS2017,TMG2015,SRB2020}. As non-zero marginal perturbations can be linked with convex decompositions of the state, they have a natural connection to the entanglement of formation \cite{C2021}.

\nocite{MATLAB,qetlab,cvx}

\section{Acknowledgements}
This work was supported, in
part, by the Quantum Valley Lower Saxony (QVLS), the
DFG through SFB 1227 (DQ-mat), the RTG 1991, and
funded by the Deutsche Forschungsgemeinschaft (DFG,
German Research Foundation) under Germany’s Excellence Strategy EXC-2123 QuantumFrontiers 390837967.


\end{document}


\title{Extremal Steering Assemblages - Supplementary Material}
	\date{\today}
	
	\author{Thomas \surname{Cope}}
	\email{thomas.cope@itp.uni-hannover.de}
	\address{Institut f{\"u}r Theoretische Physik, Leibniz Universit{\"a}t Hannover, Appelstr. 2, 30167 Hannover, Germany}
	\author{Tobias J. \surname{Osborne}}
	\email{tobias.osborne@itp.uni-hannover.de}
	\address{Institut f{\"u}r Theoretische Physik, Leibniz Universit{\"a}t Hannover, Appelstr. 2, 30167 Hannover, Germany}
	
	\begin{abstract}
	
	\end{abstract}
\maketitle

	\section{An algorithm to decompose into extremal assemblages}
	Here we present an algorithm to decompose an arbitrary assemblage into extremal ones.  Our algorithm takes in the inputs given in table \ref{Table1}, and returns a partial list of extremal assemblages, along with their weight in the final decomposition, which is complete when the recursion ends. An implementation of this algorithm in Matlab can also be found in the supplementary material. 
	
	\begin{table}
		
		\begin{center}
			\begin{tabular}{c|c}
				$\Sigma$ & The current list of extremal assemblages\\
				\vspace{\baselineskip}\\
				$P_{\Sigma}$ & The current list of weights of each extremal assemblage\\
				\vspace{\baselineskip}\\
				$\bs{\sigma}$ & The current assemblage being considered\\
				\vspace{\baselineskip}\\
				$p$ & The weight of the current assemblage being considered\\
				\vspace{\baselineskip}\\
				$x$ & The input being checked for valid perturbations\\
				\vspace{\baselineskip}\\
				$\epsilon$ & The tolerance below which values are assumed $=0$
			\end{tabular}
			\caption{The variables fed into our assemblage decomposing algorithm.}\label{Table1}
		\end{center}
	\end{table}
	
	\begin{algorithm}[H]
		\caption{Recursive Decomposition}
		\label{Recursive Algorithm}
		\begin{algorithmic}[1]
			\State First call the algorithm with inputs $\Sigma=\{\}, P_{\Sigma}=\{\}, p=1, x=R$
			\Function{Apply Recursion}{$\Sigma,P_{\Sigma},\bs{\sigma},p,x,\epsilon$}\\
			\hspace*{\algorithmicindent} \textbf{Outputs}:\\
			\hspace*{\algorithmicindent} $\Sigma$: The updated list of extremal assemblages\\
			\hspace*{\algorithmicindent} $P_{\Sigma}$: The updated list of weights of each extremal assemblage\\
			\State Define NumberInputs \label{CommStart}
			\State Define NumberOutputs
			\State	Define Dimensions\label{CommEnd} \Comment{\ref{CommStart}-\ref{CommEnd} can be taken from the dimensions of the array $\bs{\sigma}$}
			\State NumberOperators $\leftarrow0$
			\State  OperatorList $\leftarrow \{\}$
			\State  OperatorTrack $\leftarrow\{\}$
			\If{$x>0$} \Comment{We reserve $x=0$ to check non-zero marginal perturbations}
			\For{$a=1:$ NumberOutputs}\label{OutputStart}
			\State $\mathbf{d}=\mathrm{eigs}(\bs{\sigma}_{a|x}$)\Comment{Take \textbf{d} in descending order}
			\State Define matrix $V$ where  $\bs{\sigma}_{a|x}\mathbf{v}_r=d_r\mathbf{v}_r$
			\For{r=1:Dimension}
			\If{$d_r>\epsilon$}
			\State insert $\ket{v_r}\bra{v_r}$ into Operator List
			\State insert $(a,x)$ into OperatorTrack
			\State NumberOperators $\leftarrow$ NumberOperators$+1$
			\For{$r_2=1:r-1$}
			\State insert $1/\sqrt{2}(\ket{v_r}\bra{v_{r_2}}+\ket{v_{r_2}}\bra{v_{r}}$ into OperatorList
			\State insert $(a,x)$ to OperatorTrack
			\State NumberOperators $\leftarrow$ NumberOperators$+1$.
			\State insert $1/(\sqrt{2}i)(\ket{v_r}\bra{v_{r_2}}-\ket{v_{r_2}}\bra{v_{r}}$ into OperatorList
			\State insert $(a,x)$ to OperatorTrack
			\State NumberOperators $\leftarrow$ NumberOperators$+1$.
			\EndFor
			\EndIf
			\EndFor
			\EndFor \label{OutputEnd} 
			\State Create $M$ whose $i$\textsuperscript{th} column is the $i$\textsuperscript{th} operator in OperatorList \Comment{one has to reshape matrices to vectors}
			\State Find the nullspace of $M$ \Comment{each vector is a valid perturbation}
			\If{null(M)$ =\{0\}$}
			\State Apply Recursion($\Sigma,P_{\Sigma},\bs{\sigma},p,x-1,\epsilon$) \comment{we can move on to the next input}
			\Else
			\State Choose perturbation $\sum_i c_i O_{i}=0$, where $O_{i}\in$ OperatorList \label{PerturbationStart} \Comment{taken from the nullspace}
			\State Construct perturbation $\mathbf{D}$, where $D_{a|x}=\sum_{\text{OperatorTrack}(i)=(a,x)} O_{i}$
			\State Maximise $w_+$ such that $\sigma_{a|x} + w_+D_{a|x} \geq 0, \forall a,x$ \Comment{done via semidefinite programming}
			\State Maximise $w_-$ such that $\sigma_{a|x} - w_-D_{a|x} \geq 0, \forall a,x$
			\State Calculate relative weights $p_+ = w_-/(w_+ + w_-)$, $p_- = w_+/(w_+ + w_-)$
			\State Apply Recursion($\Sigma,P_{\Sigma},\bs{\sigma}+w_+\mathbf{D},p*p_+,x,\epsilon$) \Comment{We apply the function recursively}
			\State Apply Recursion($\Sigma,P_{\Sigma},\bs{\sigma}-w_-\mathbf{D},p*p_-,x,\epsilon$)\label{PerturbationEnd}
			\EndIf
			\Else \Comment{When $x=0$, we look for marginal-changing perturbations}
			\For{x=1:NumberInputs}
			\State Run lines \ref{OutputStart}-\ref{OutputEnd}
			\EndFor
			\State Construct a matrix $M$ such that $M\boldsymbol{u}$ constrains:\\
			\hspace*{\algorithmicindent}\hspace*{\algorithmicindent} $\sum_{\text{OperatorTrack}(i)=(a,1)} u_iO_i = \sum_{\text{OperatorTrack}(j)=(a,x)} u_jO_j$,\;\;$\sum_{\text{OperatorTrack}(i)=(a,1),\mathrm{Tr}(O_i)=1} v_i = 0$
			\If{nullspace(M)$ =\{0\}$}
			\State Add $\bs{\sigma}$ to $\Sigma$, and $p$ to $P_{\Sigma}$
			\Else
			\State Run lines \ref{PerturbationStart}-\ref{PerturbationEnd}
			\EndIf	
			\EndIf
			\EndFunction
		\end{algorithmic}
	\end{algorithm}